\documentclass[manuscript]{aastex}

\slugcomment{ApJ Letters}

\shorttitle{Comet C/2019 Q4}
\shortauthors{Jewitt and Luu}

\begin{document}


\title{Initial Characterization of Interstellar Comet 2I/2019 Q4 (Borisov)}


\author{David Jewitt$^{1,2}$ and Jane Luu$^{3}$}
\affil{$^1$ Department of Earth, Planetary and Space Sciences,
UCLA, 
595 Charles Young Drive East, 
Los Angeles, CA 90095-1567\\
$^2$ Dept.~of Physics and Astronomy,
UCLA, 
430 Portola Plaza, Box 951547,
Los Angeles, CA 90095-1547\\
$^3$ Centre for Earth Evolution and Dynamics, University of Oslo, Postboks 1028 Blindern, NO-0315 Oslo, Norway}


\email{jewitt@ucla.edu}

\begin{abstract}
We present initial observations of the interstellar body  2I/(2019 Q4) Borisov taken to determine its nature prior to the perihelion in 2019 December.  Images from the Nordic Optical Telescope show a prominent, morphologically stable dust coma and tail.  The dust cross-section within 15,000 km of the nucleus averages 130 km$^2$ (assuming geometric albedo 0.1) and increases by about 1\% per day. If sustained, this rate indicates that the comet has been active for $\sim$100 days prior to the observations.  Cometary activity thus started in 2019 June, at which time C/Borisov was at $\sim$4.5 AU from the Sun, a typical distance for the onset of water ice sublimation in comets.   The dust optical colors, B-V = 0.80$\pm$0.05, V-R = 0.47$\pm$0.03 and R-I = 0.49$\pm$0.05 are identical to those of a sample of (solar system) long-period comets. The colors are similar to those of 1I/(2017 U1) 'Oumuamua, indicating a lack of the ultrared matter that is common in the Kuiper belt, on both interstellar objects.  The effective size of the dust particles is estimated as $\overline{a}$ = 100 $\mu$m, based on the length of the dust tail and the 100 day lifetime.  With this size, the ejected dust mass is of order  1.3$\times10^7$ kg and the current dust mass loss rate $\sim$ 2 kg s$^{-1}$.  We set an upper limit to the nucleus radius using photometry at $r_n \le$ 3.8 km (again for albedo 0.1) and we use a statistical argument to show that the nucleus must be much smaller, likely a few hundred meters in radius.  
\end{abstract}


\section{INTRODUCTION}
Object 2I/(2019 Q4) Borisov (hereafter ``Q4'') was discovered by  G. Borisov of the Moscow State University on UT 2019 August 30 and publicly announced on September 11 (Borisov 2019).  It is the second known interstellar object in the solar system, after 1I/(2017 U1) 'Oumuamua, and the first interstellar comet.    Discovered at only 38\degr~solar elongation, Q4 represents both a triumph of small-telescope astronomy and a challenge for observers using large telescopes, few of which can be operated  at such small angles from the Sun.  In this report, we describe initial observations from the 2.56 m diameter Nordic Optical Telescope (NOT), designed to provide a first characterization of the object.  

\section{OBSERVATIONS}
The small elongation of Q4 and the requirement that optical observations be taken against a dark sky forced us to observe at low elevations immediately before sunrise.  The NOT,  located at 2400 m altitude in the Canary Islands, can take useful data at elevations as small as 6.4\degr~(airmass 9).  
    On UT 2019 September 13 and 14, we employed the 1024$\times$1024 pixel StanCam, with 0.176\arcsec~pixels giving a 3.0\arcmin$\times$3.0\arcmin~field of view.  On September 15, 18, 26 and October 04 we used the 2048$\times$2048 pixel ALFOSC camera, which has a 6.5\arcmin$\times$6.5\arcmin~field of view with 0.214\arcsec~pixels.  The telescope was tracked  non-sidereally to follow the motion of the comet (approximate rates 60\arcsec~hour$^{-1}$ East and 50 \arcsec~hour$^{-1}$ South).  We began observing at airmasses as high as $\sim$8 in observing windows that were soon truncated by morning twilight.  Furthermore, owing to the urgency of the observations our initial data were taken in the presence of scattered light from the full Moon.  Both the airmass and the Moon phase improved, however, and the later observations were possible at more modest airmasses, $\sim$2 to 3, and against a dark sky.    A journal of observations is given in Table (\ref{geometry}).

We used   broadband BVRI filters approximating the Bessel (1995) system to measure  Q4.  The central wavelengths, $\lambda_c$, and full widths at half maximum (FWHM) of the ALFOSC filters in the form Filter($\lambda_c$, FWHM) are B(4400,1000), V(5300,800), R(6500,1300) and $i_{int}$(7970,1570), with all wavelengths expressed in \AA.  The StanCam V filter is slightly different, V(5430,1030).  Flat fields were constructed after debiasing using nightly images of the illuminated interior of the observatory dome.  The data were photometrically calibrated both with respect to field stars in the Sloan DR14 data release (Blanton et al.~2017) and through observations of Landolt (1992) photometric standard stars.   Use of the Sloan field stars entails no airmass correction, but necessitates a transformation from the Sloan magnitude system to Bessel magnitudes using the relations given by Jordi et al.~(2006).  To use the Landolt stars, which were necessarily observed at airmasses different from those of the comet, we measured and applied extinction coefficients  of $k_B$ = 0.23, $k_V$ = 0.14 and $k_R$ = 0.11 magnitudes per airmass.  We did not measure $k_I$ but instead assume $k_I$ = 0.06 magnitudes per airmass.

%

\section{DISCUSSION}
\begin{figure}
\plotone{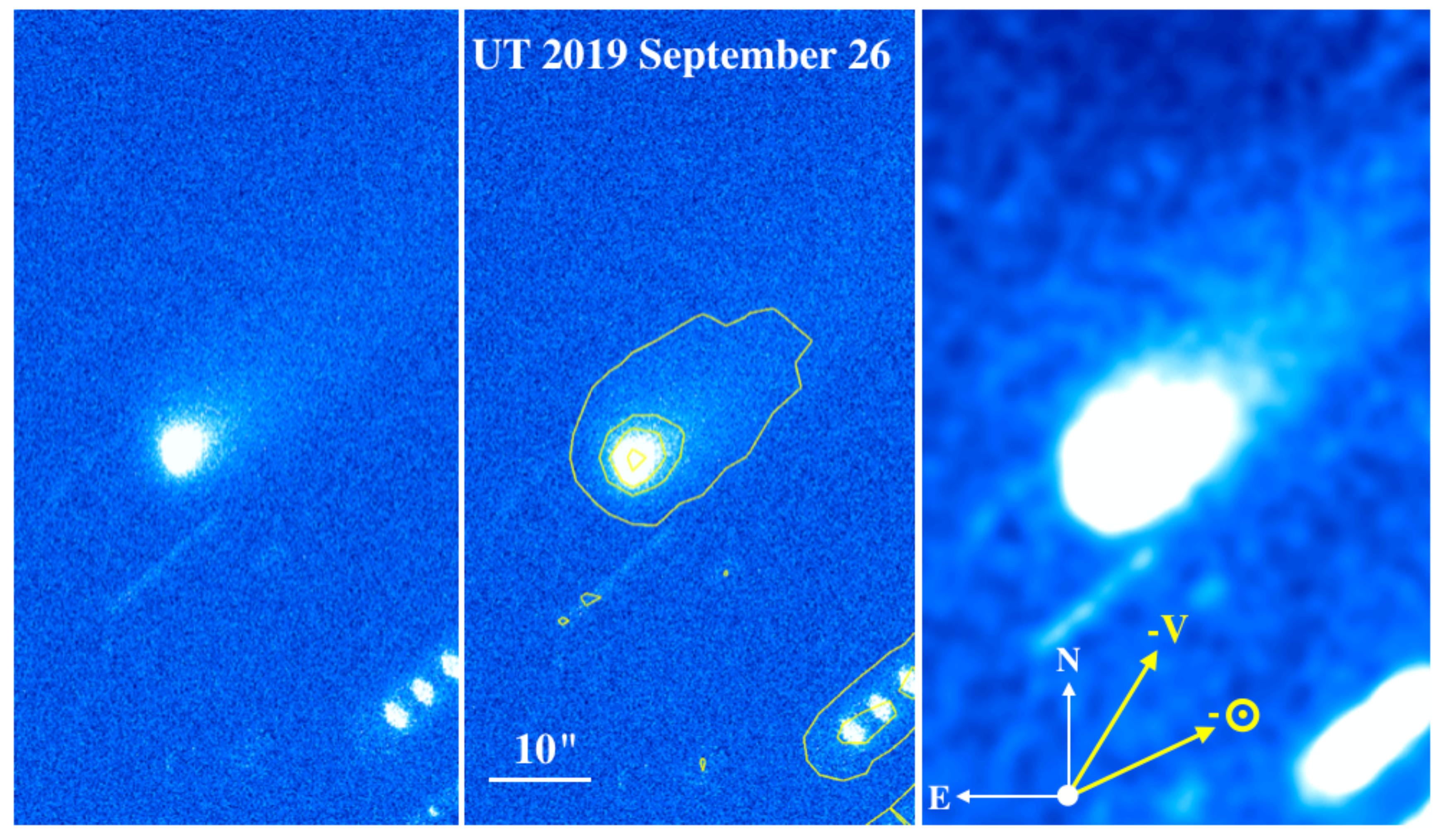}
\caption{Composite of eight  images of Q4, each of 180 s duration, taken through the R filter on UT 2019 September 26.  The three panels show (left) the raw composite, (middle) added contours to highlight the coma and (right) a spatially smoothed version, to show the faintest dust.  White arrows show the directions of North and East, while yellow arrows marked $-\odot$ and $-V$ show the projected anti-solar and anti-velocity vectors.  A 10\arcsec~(2.3$\times10^4$ km) scale bar is shown. \label{september26}}
\end{figure}

\noindent \textbf{Morphology:} Except for differences in the sensitivity to low surface brightness material caused by nightly variations in the sky brightness, the appearance of Q4 did not change between the different nights of observation (Table \ref{geometry}). Figure (\ref{september26}) shows a representative R-band  image composite from UT 2019 September 26 (fraction of Moon illuminated $\sim$10\%) formed by aligning and combining eight images each of 180 s duration.  The left panel shows the unadorned image, the middle panel adds contours to highlight the tail and the right panel has been smoothed by convolution with a gaussian function having FWHM = 1\arcsec~to emphasize faint structure.  Vectors $-V$ and $-\odot$ show the projected negative orbital velocity and the projected anti-solar direction (see also Table \ref{geometry}). The comet is clearly non-stellar, and shows an extensive dust tail to the north west, approximately bounded by the projected orbit and anti-solar vectors, as is a characteristic of dust tails.  The visible portion of the tail is limited to about 60\arcsec~in length by sky noise and field structure from trailed field stars and galaxies.  This corresponds to a sky-plane length $L = 1.4\times10^8$ m.    If the tail is in fact anti-solar, then its true length is given by $L_0 = L/sin(\alpha)$ which, with $\alpha$ = 17\degr~(Table \ref{geometry}), gives $L_0 = 4.8\times10^8$ m.  

\begin{figure}
\epsscale{.95}
\plotone{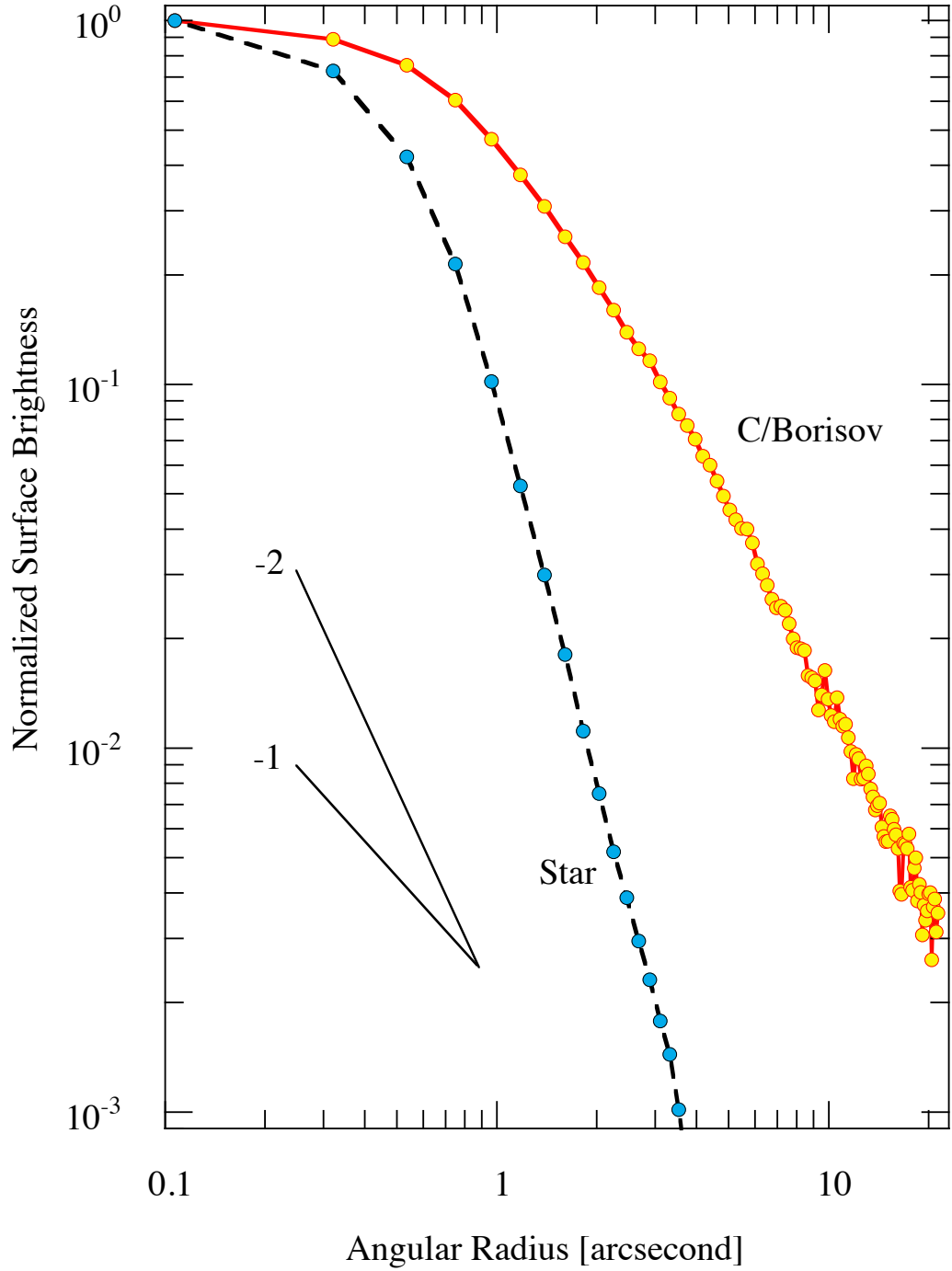}
\caption{Surface brightness profiles of Q4 and a field star from UT 2019 September 26, in the R filter.   Straight lines show surface brightess gradients $m$ = -1 and $m$ = -2, as marked.  \label{profile_sep26}}
\end{figure}

The surface brightness profile of Q4 is compared with that of a field star in Figure (\ref{profile_sep26}).  Both profiles were computed by averaging the signal within a set of concentric circular apertures centered on the optocenters of each object.  Sky subtraction was determined from the median signal within a concentric annulus having inner and outer radii 200 pixels (42.8\arcsec) and 107.0\arcsec, respectively.  We experimented with the radii of the sky annulus, finding no significant effect on the profile over the region measured.  The figure shows the extended nature of Q4.  The central region of the profile is strongly affected by the point-spread function of the data and we do not attempt to model it here.  We fitted a power law  to the surface brightness profile over the radius range 5\arcsec~to 22\arcsec, finding $\Sigma(\theta) \propto \theta^{m}$, where $\Sigma(\theta)$ is the normalized surface brightness at radius $\theta$ and index $m$ = -1.85$\pm$0.02.  This value is steep compared to $m$ = -1, as expected of an isotropic coma in steady state, and also steeper than the value $m$ = -3/2 resulting from the action of radiation pressure on an otherwise steady-state coma (Jewitt and Meech 1987).  Interpretation of this profile is deferred to the aquisition of more data on Q4 as it rounds perihelion.  We merely note that steeper profiles can result from fading grains or, more plausibly, a dust production rate rising with time.  

\begin{figure}
\epsscale{.95}
\plotone{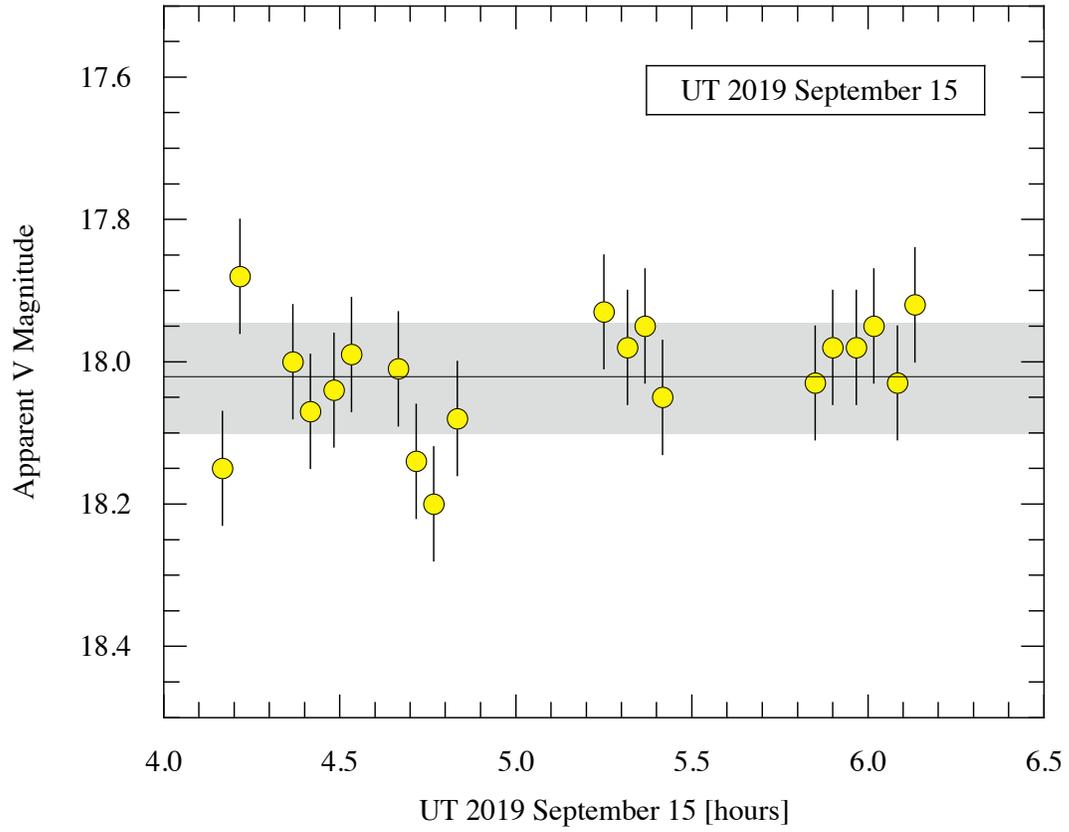}
\caption{Lightcurve on UT 2019 September 15 measured in the V filter within a projected aperture 15,000 km in radius. The horizontal line shows the mean value, V = 18.02.  \label{lightcurve}}
\end{figure}

\noindent \textbf{Photometry:} We measured the brightness of Q4 in each image within circular apertures having projected radii 7,500 km and 15,000 km (roughly 3\arcsec~and 6\arcsec, respectively, although varying with the geocentric distance to Q4).  Sky subtraction was obtained using the median level within a contiguous, concentric annulus of width 10.7\arcsec.   The  difference between Q4 and the field stars was used to negate the effects of changing atmospheric extinction.  After this correction, no convincing photometric variability was detected within each night.  As an example,  Figure (\ref{lightcurve}) shows V-filter photometry from a $\sim$2 hour  timespan on UT 2019 September 15, during which time the comet rose from airmass 6.5 to 1.9.  Representative error bars of $\pm$0.08 magnitudes are included.  The mean and standard error on the mean of the plotted data are V = 18.02$\pm$0.03 ($n$ = 20 measurements), with no evidence for a systematic trend in the magnitude over this period.   Deviations on timescales $\sim$30 minutes are likely related to seeing and guiding fluctuations given the high airmass of these observations, especially near the beginning of the observing window (UT 4 to 5 hours).  This photometric invariance is a natural result of coma dilution within the photometric aperture (Jewitt 1991).  This occurs when the  timescale for particles to cross the aperture, $\tau_{cross}$, is comparable to or longer than the timescale for variation of the source.  For example, small dust particles well-coupled to the outflowing gas would leave the nucleus with a speed comparable to the speed of sound in gas at the local blackbody temperature ($V_s \sim$ 0.4 km s$^{-1}$ for gaseous H$_2$O with $T_{BB}$ = 168 K at 2.74 AU).   Then, with aperture radius $\ell$ = 15,000 km, we find  $\tau_{cross} = \ell/V_s \sim 3.8\times10^4$ s (about 10 hours), and larger (slower) particles will take a longer time.  Photometric variations on timescales $\lesssim$ 10 hours must necessarily be damped by aperture-averaging.  For comparison, 1I/(2017 U1) 'Oumuamua was devoid of coma and showed an extreme lightcurve with period $\sim$8 hours and a range $\sim$2.5 magnitudes (Meech et al.~2017), indicating an axis ratio of $\sim$5:1 (Bannister et al.~2017, Drahus et al.~2018). Coma dilution would render such a lightcurve in Q4 invisible.

In order to search for variations on longer timescales, we compared the averaged photometry from each night of observation.  The observing geometry changes significantly between nights (Table \ref{geometry}) so we compared absolute magnitudes, $H$,  computed using

\begin{equation}
H = V - 5\log_{10}(r_h \Delta) - f(\alpha)
\label{H}
\end{equation}

\noindent where $V$ is the apparent magnitude and $f(\alpha)$ is the phase function.   The backscattering phase functions of comets are in general poorly known and that of Q4 is completely unmeasured.  We used $f(\alpha) = 0.04\alpha$, which gives the ratio of scattered fluxes at 0\degr~phase and 15\degr~phase as $B$ = 1.7, comparable to values $B \sim$ 2 measured in 67P/Churyumov-Gerasimenko (Bertini et al.~2019).  

The absolute magnitude is further related to the effective scattering cross-section, $C_e$ [km$^2$], by 
\begin{equation}
C_e = \frac{1.5\times 10^6}{p_V} 10^{-0.4 H}
\label{area}
\end{equation}

\noindent where $p_V$ is the geometric albedo.  We assume $p_V$ = 0.1, as appropriate for cometary dust (Zubko et al.~2017).  A much higher albedo could apply if the coma grains were icy, but the spectroscopic non-detection of water ice absorption bands reported by Yang et al.~(2019) suggests that this is not the case.  The nightly apparent and absolute magnitudes and the scattering cross-sections are listed  in Table (\ref{photometry}). 

\begin{figure}
\epsscale{.95}
\plotone{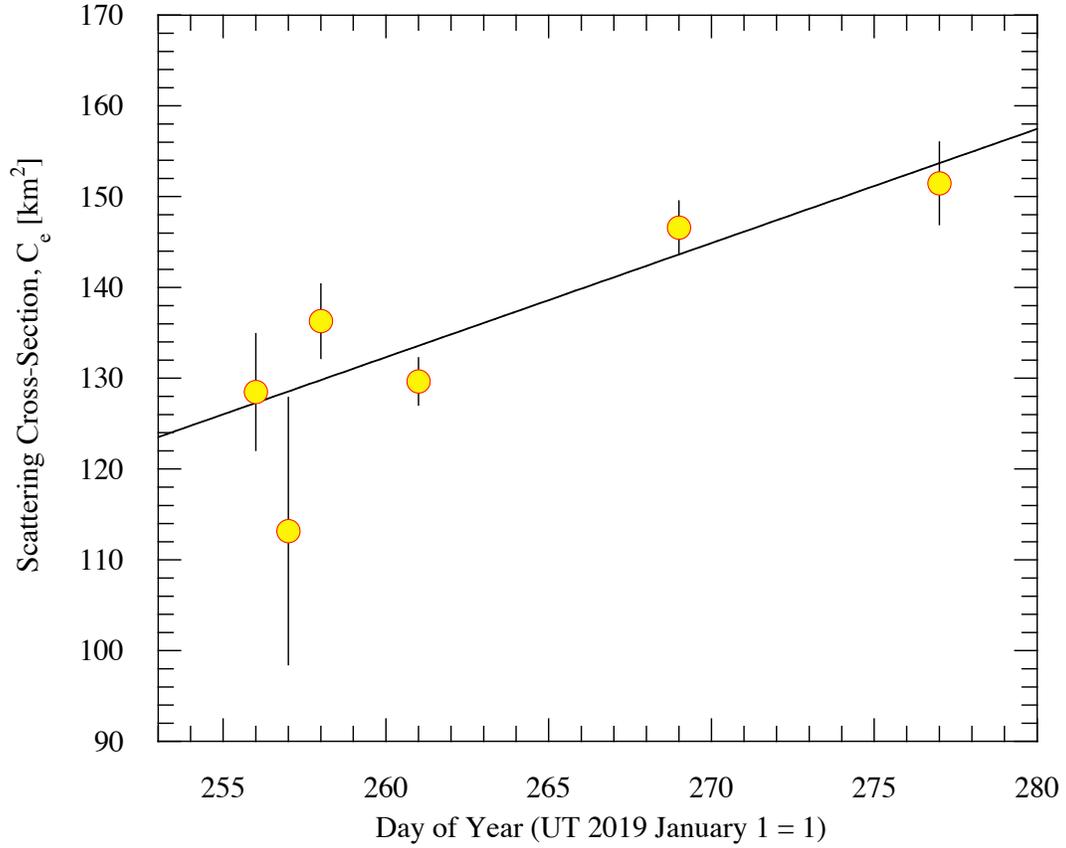}
\caption{Scattering cross-section within a circular aperture 1.5$\times10^4$ km in radius, as a function of time, expressed as Day of Year (DOY = 1 on UT 2019 January 1).   The line shows a linear, least-squares fit having gradient 1.26$\pm$0.25 km$^2$ day$^{-1}$.  Data from Table (\ref{photometry}). \label{h_vs_time}}
\end{figure}

The average cross-section within the 15,000 km aperture is $C_e \sim$ 130 km$^2$, with a slight dependence on time.    Figure (\ref{h_vs_time}) shows $C_e$ vs.~time together with a weighted linear, least-squares fit to the data having best-fit gradient $dC_e/dt = 1.26\pm0.25$ km$^2$ day$^{-1}$ (i.e.~about 1\% day$^{-1}$).  The magnitude of $dC_e/dt$ is influenced by the adopted phase function but, since the range of phase angles in our data is small (Table \ref{geometry}) the effect is modest.  Phase functions in the range $f(\alpha)$ = 0.03 to 0.05 magnitudes degree$^{-1}$ change the gradient by an amount smaller than the statistical error.  Extrapolation of the data gives $C_e$ = 0 km$^2$ about 100 days before  the first observation on September 13, corresponding to DOY 156 (UT 2019 June 5, when Q4 was at $r_H$ = 4.5 AU).   We possess no proof that such an extrapolation is justified, but it is noteworthy that the inferred turn-on distance matches the $r_H \sim$ 4 AU to 5 AU critical distance at which water ice sublimation begins in comets.  

The optical  cross-section is almost entirely carried by dust and we can use it  only to derive  upper limits to the size of the nucleus.  A crude upper limit is given by $r_n = (C_e/\pi)^{1/2}$ with $C_e$ = 130 km$^2$.  Substituting gives $r_n \le 6.4$ km, again assuming $p_V$ = 0.1.    We sought a  stronger  limit  using photometry from smaller, less dust-contaminated apertures. By experimentation, we found that photometry within apertures of radius $<$2.1\arcsec~(10 ALFOSC pixels, or about 4800 km at the distance of Q4) was unduly sensitive to variations in the image point spread function caused by a combination of atmospheric turbulence and telescope tracking.  Photometry within a 2.1\arcsec~radius aperture on UT 2019 September 26, with background subtraction of the coma from a contiguous annulus of outer radius 4.2\arcsec, gives V = 19.03$\pm$0.03.  This corresponds through Equation (\ref{H}) to $H$ = 13.82 and through Equation (\ref{area}) to $C_e$ = 45 km$^2$ and $r_n < 3.8$ km.   We emphasize that this is still a strong upper limit to $r_n$ because of dust contamination in the photometry aperture.  More stringent observational constraints on the nucleus await the acquisition of high angular resolution data and/or the cessation of activity as Q4 recedes from the Sun, post-perihelion.

The mass of dust, $M$, and its cross-section, $C_e$ are related by  $M \sim 4  \rho \overline{a} C_e/3$, where $\overline{a}$ is the mean dust particle radius and  $\rho$ is the particle density.  We take $\rho$ = 10$^3$ kg m$^{-3}$ as the nominal density.  The mean particle radius is obtained from the tail length, mentioned above as $L_0 = 4.8\times10^8$ m.    We assume that these particles were ejected from the nucleus a time $t \sim$ 100 days ($\sim9\times10^6$ s) ago and that their deflection into a tail is the result of solar radiation pressure acceleration.  We write the radiation pressure acceleration as $\beta g_{\odot}(1) r_H^{-2}$, where $\beta$ is a (dimensionless) function of the particle properties and $g_{\odot}(1)$ = 0.006 m s$^{-2}$ is the gravitational acceleration towards the Sun at $r_H$ = 1 AU.  Neglecting the variation of $r_H$ over the 100 day flight time, and neglecting their initial velocity, we estimate $\beta$ from

\begin{equation}
\beta \sim \frac{2L_0 r_H^2}{g_{\odot} t^2}
\end{equation}

\noindent with $r_H$ expressed in AU. We set $r_H$ =   2.580 AU (September 26) to find $\beta \sim$ 0.01.  In dielectric spheres, $\beta$ is approximately equal to the inverse particle radius expressed in microns (Bohren and Huffman 1983).  Therefore, $\beta \sim$ 0.01 implies effective particle size $\overline{a}$ = 100 $\mu$m.  Strictly, this estimate applies to particles displaced to the end of the visible tail and particles closer to the nucleus could be much larger.  We preliminarily take $\overline{a}$ = 100 $\mu$m as the nominal particle radius, pending more accurate determinations.

Substituting, we find dust mass $M \sim 1.3\times10^7$ kg.  The rate of production of dust is $dM/dt = 4 \rho \overline{a} (dC_e/dt)/3 \sim 2\times10^5$ kg day$^{-1}$ (2 kg s$^{-1}$).    For comparison, the only other currently available constraint on the mass loss rate  is from a  reported CN production rate $Q_{CN} \sim  (3.7\pm0.4)\times 10^{24}$ s$^{-1}$ (0.2 kg s$^{-1}$; Fitzsimmons et al.~2019).  In solar system comets, the ratio of the water to CN production rates varies about an average value $Q_{H2O}/Q_{CN}$ = 360 (A'Hearn et al.~1995).  If this ratio applies to Q4, then we infer a mass loss rate in water of $\dot{M} \sim$ 60 kg s$^{-1}$, more than an order of magnitude larger than the  production rate in dust.  The equilibrium mass sublimation flux at 2.7 AU for an absorbing water ice surface oriented perpendicular to the Sun direction is $f_s = 4\times10^{-5}$ kg m$^{-2}$ s$^{-1}$, which could be supplied by a patch of area $A = \dot{M}/f_s \sim 1.5\times10^6$ m$^2$ (1.5 km$^2$).  This is equal to the surface area of a sphere of radius $r_N$ = 0.35 km and sets a lower bound to the  radius of the nucleus, assuming that nucleus sublimation is the only gas source.   By similar arguments, Fitzsimmons et al.~(2019) found an upper limit to the radius $r_n <$ 8 km and a preferred range 0.7 $\le r_n \le$ 3.3 km.

\begin{figure}
\epsscale{.80}
\plotone{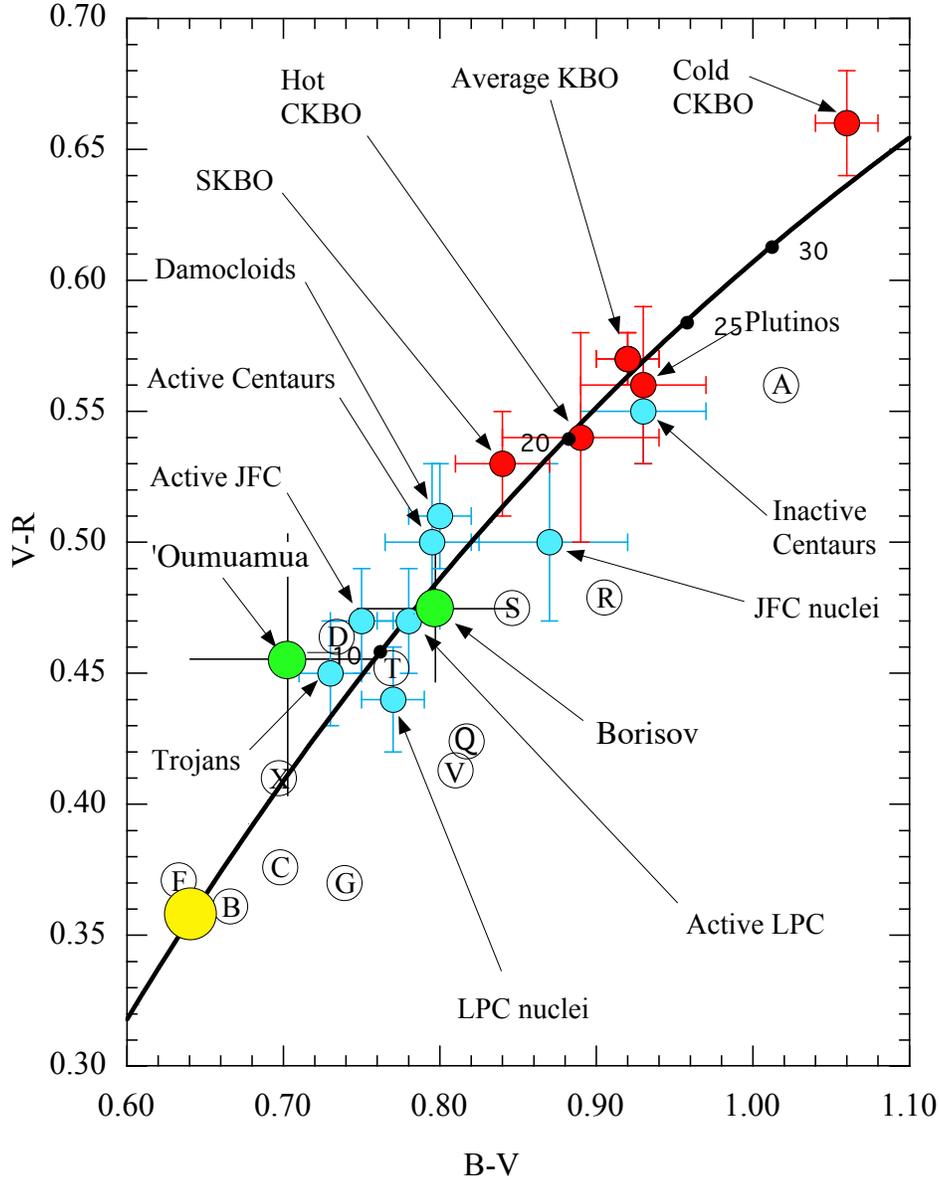}
\caption{The B-V vs.~V-R color plane comparing 1I/(2017 U1) 'Oumuamua and Q4 with  solar system objects.  Red circles indicate sub-types of Kuiper belt object (hot and cold classical KBOs, 3:2 resonant ``Plutinos'' and scattered KBOs are distinguished), blue circles indicate the Centaurs and the nuclei and comae of both short and long-period comets, as labeled, together with the Jupiter Trojans.  All data  from Jewitt (2015).  Letters show the positions of main-belt asteroid spectral classes according to Dandy et al.~(2003).   \label{color_color}}
\end{figure}

\noindent \textbf{Colors:} The mean colors  listed in Table (\ref{colors}) show no dependence on the aperture radius between 3\arcsec~and 6\arcsec.  Our measurement of V-R can be compared with an independent determination, g-r = 0.63$\pm$0.02 (Guzik et al.~2019, who used a 2\arcsec~radius aperture) which, when transformed using the relations of Jordi et al.~(2006), gives V-R = 0.49$\pm$0.02.  The agreement is excellent.  The Table also lists the colors of the Sun, of 1I/(2017 U1) 'Oumuamua (Jewitt et al.~2017) and the mean color of the long period comets (Jewitt 2015). Figure (\ref{color_color}) shows the B-V vs.~V-R color plane for these and other solar system objects, modified from Jewitt (2015).  The optical colors of Q4 are  redder than the Sun (the reflectivity gradient across the BR region of the spectrum is $S' \sim$ 4\% per 1000\AA) but closely match the mean colors of (solar system) long-period comets.  The colors of Q4 are similar to those of 1I/(2017 U1) 'Oumuamua within the uncertainties of measurement.  Neither interstellar object shows evidence for the ultrared matter ($S' \ge$ 25\%/1000 \AA) which is a prominent feature of many Kuiper belt objects (Jewitt 2002, 2015).  The lack of ultrared matter (likely to consist of complex irradiated organics, c.f.~Cruikshank et al.~1998, Dalle Ore et al.~2015) mirrors its absence in the solar system at distances $\lesssim$10 AU.  As in the active comets of the solar system, we surmise that the particles ejected into the coma and tail of Q4 are derived from beneath a pre-existing ultrared mantle of cosmic-ray irradiated material.   The ultrared matter is either thermodynamically unstable in the inner solar system as a result of the elevated temperatures or perhaps ejected or buried by fallback debris (Jewitt 2002).

\noindent \textbf{Statistics of Interstellar Objects:}  The discovery of Q4 two years after 1I/'Oumuamua exactly matches a published prediction of the discovery rate, namely $S \sim$ 0.5 to 1 year$^{-1}$ (Jewitt et al.~2017).  While this match at first appears gratifying,  it is difficult to quantitatively compare the discoveries of the two interstellar objects given that they have such different physical properties and that they were identified in surveys having very different sky coverage, depth and cadence parameters.  Moreover, the visibility of Q4 is enhanced by a coma, without which the object would likely not have been noticed, whereas 'Oumuamua appeared persistently unresolved.  

Based on 'Oumuamua alone the number density of similarly-sized interstellar objects was estimated as $N_1 \sim 0.1$ AU$^{-3}$ and the differential power-law size distribution as $r_n^{-q}$ with $q \ge$ 4 (Jewitt et al.~2017).  Objects larger than 'Oumuamua should accordingly be very rare.   The latter was elongated in shape but  had an effective radius variously estimated (in order of size) as $r_O$ = 45 to 90 m (Knight et al.~2017),  $\sim$ 55 m (Jewitt et al.~2017), 75 to 79 m (Drahus et al.~2018),  $\le$100 m (Bannister et al.~2017), 102 m (Meech et al.~2017) and $\le$130 m (Bolin et al.~2018).  For the sake of argument, we take $r_O$ = 0.1 km as the effective radius of 'Oumuamua.  Given $q$ = 4, the cumulative number of interstellar objects larger than radius $r_n$ (km) and inside a sphere of  radius $r_H$ (AU) is just 

\begin{equation}
N(r_n) = \frac{4\pi r_H^3}{3}N_1 \left[\frac{r_O}{r_n}\right]^3.   
\label{N}
\end{equation}

\noindent Q4 was discovered at $r_H$ = 3 AU.  Setting $r_n$ = 3.8 km (the upper limit to the radius set by our photometry) in Equation (\ref{N}), we find $N(3.8)$ = 2$\times10^{-4}$.  We thus consider it very unlikely that Q4 could be as large as our photometric limit allows.   In fact,  the nucleus is very unlikely to be larger than 1 km in radius and is most probably just a few hundred meters.  For example, Equation (\ref{N}) gives $N$ = 1 for $r_n \sim$ 0.2 km and $N$ = 0.1 for $r_n$ = 0.5 km.   A subsequent refinement of our number density estimate of 'Oumuamua-sized bodies (from $N_1$ = 0.1 AU$^{-3}$ to $\sim$0.2 AU$^{-3}$) by Do et al.~(2018) does not materially change this conclusion.

Published upper limits to the number density of interstellar objects  fall in the range $N_1(1) = 10^{-3}$ to 10$^{-5}$ AU$^{-3}$, as summarized by Engelhardt et al.~(2017).  Engelhardt's own best value is  $N_1(1) < 10^{-4}$ AU$^{-3}$, based on involved modeling of three sky surveys each giving zero detections.  With $N_1(1) < 10^{-4}$ AU$^{-3}$, the number of 1 km radius objects expected within $r_H$ = 3 AU of the Sun is $N < 10^{-2}$, again showing that the nucleus of Q4 is likely to be a sub-kilometer body.  


Inbound long period comet C/2019 J2 recently disintegrated when 1.9 AU from the Sun (Jewitt and Luu 2019), a distance essentially equal to the $q$ = 2.0 AU perihelion distance of Q4.  Disintegration is a common yet poorly quantified property of  comets, especially those with small nuclei and perihelia (c.f.~Sekanina and Kracht 2018).  The nucleus of C/2019 J2, like that of Q4, had a radius of only a few hundred meters and was probably  rotationally disrupted  by torques from anisotropic outgassing (Jewitt and Luu 2019).  Continued observations are encouraged to determine whether Q4 might undergo the same fate and, if so, whether it might leave behind a low activity remnant resembling 'Oumuamua.

\clearpage 

\section{SUMMARY}
We present observations of interstellar comet 2I/(2019 Q4) Borisov taken at small solar elongation with the 2.56 m Nordic Optical Telescope.

\begin{enumerate}

\item The comet is actively emitting dust, as evidenced by a tail of sky-plane length $> 1.4\times10^8$ m, and by progressive brightening of the coma at about 1\% day$^{-1}$.  

\item We infer that strong activity began near heliocentric distance 4.5 AU as the comet entered the water ice sublimation zone.  The effective particle radius is $\overline{a} \sim$ 100 $\mu$m, the coma mass $M \sim 1.3\times10^7$ kg, and the production rate in dust $dM/dt \sim$ 2 kg s$^{-1}$.

\item An observational upper limit to the nucleus radius is set at $r_n < 3.8$ km (albedo $p_V$ = 0.1 assumed).  
However, based on statistical considerations, we find that the nucleus must be much smaller, and is likely just a few hundred meters in radius.

\item The optical colors  B-V = 0.80$\pm$0.05, V-R = 0.47$\pm$0.03, R-I = 0.49$\pm$0.05, are slightly redder than the Sun, similar to 1I/(2017 U1) 'Oumuamua and identical within the uncertainties of measurement to the mean color measured for the dust comae of long-period comets.  Like active solar system objects,  both interstellar objects lack ultrared matter.

\end{enumerate}

\acknowledgments

We thank Yoonyoung Kim for comments on the manuscript, David Mkrtichian for a discussion in Kunming, Anlaug Amanda Djupvik for help with the observations and Thomas Augusteijn for allocating time to this project.

{\it Facilities:}  \facility{NOT}.

\clearpage

\clearpage

\clearpage

\clearpage

\clearpage

\clearpage

\begin{deluxetable}{lcccrccccr}
\tablecaption{Observing Geometry 
\label{geometry}}
\tablewidth{0pt}
\tablehead{ \colhead{UT Date and Time} & Airmass\tablenotemark{a}   & DOY\tablenotemark{b} & $\Delta T_p$\tablenotemark{c} & \colhead{$r_H$\tablenotemark{d}}  & \colhead{$\Delta$\tablenotemark{e}} & \colhead{$\alpha$\tablenotemark{f}}   & \colhead{$\theta_{\odot}$\tablenotemark{g}} &   \colhead{$\theta_{-V}$\tablenotemark{h}}  & \colhead{$\delta_{\oplus}$\tablenotemark{i}}   }
\startdata

2019 Sep 13  04:12  - 06:02  & 6.49 - 1.94   & 256 &  -86     &  2.767 &  3.407  & 14.5 & 298.6   & 326.8 & -7.9 \\
2019 Sep 14  04:22 - 06:05   &  5.18 - 1.90  & 257 &  -85     &  2.752  & 3.385  & 14.7  & 298.3  & 326.9 & -8.1 \\
2019 Sep 15  04:40 - 06:08   &  3.94 - 1.85  & 258 &  -84     &  2.737  & 3.363  & 14.9  & 298.0  & 327.0 & -8.3 \\
2019 Sep 18  04:01 - 05:17   &  7.16 - 2.54  & 261 &  -81     &  2.693  & 3.297  & 15.5  & 297.3  & 327.4  & -8.9 \\
2019 Sep 26 05:03 - 06:00    & 2.59 - 1.76    & 269 & -73      &  2.580 & 3.123 & 17.1 & 295.5 & 328.4 & -10.4 \\
2019 Oct 04 05:25 - 05:49     & 2.02 - 1.75    & 277 & -65     &  2.470 & 2.950 & 18.7 & 294.0 & 329.3 & -12.1 \\

\enddata


\tablenotetext{a}{Airmass at the start and end time of observation}

\tablenotetext{b}{Day of Year, UT 2019 January 01 = 1}
\tablenotetext{c}{Number of days from perihelion (UT 2019-Dec-08 = DOY 342). Negative numbers indicate pre-perihelion observations.}
\tablenotetext{d}{Heliocentric distance, in AU}
\tablenotetext{e}{Geocentric distance, in AU}
\tablenotetext{f}{Phase angle, in degrees}
\tablenotetext{g}{Position angle of the projected anti-Solar direction, in degrees}
\tablenotetext{h}{Position angle of the projected negative heliocentric velocity vector, in degrees}
\tablenotetext{i}{Angle of Earth above the orbital plane, in degrees}

\end{deluxetable}

\clearpage

\begin{deluxetable}{lccl}
\tablecaption{Photometry with Fixed  Linear  Apertures
\label{photometry}}
\tablewidth{0pt}
\tablehead{
\colhead{UT Date}    & \colhead{V\tablenotemark{a}}   &   \colhead{$H$\tablenotemark{b}}&  \colhead{$C_e$\tablenotemark{c}}}

\startdata
September 13	& 18.12$\pm$0.05	& 12.67	& 128$\pm$6	  \\

September 14	& 18.24$\pm$0.13	& 12.81	& 113$\pm$16	  \\

September 15	& 18.02$\pm$0.03	& 12.60	& 136$\pm$4	  \\

September 18	& 18.02$\pm$0.02	& 12.66	& 130$\pm$3	  \\

September 26	& 17.74$\pm$0.02	& 12.53	& 146$\pm$3	  \\

October 04      & 17.55$\pm$0.03     & 12.49     & 151$\pm$5     \\

\enddata


\tablenotetext{a}{Apparent V-band magnitude within 15,000 km radius projected aperture}

\tablenotetext{b}{Absolute magnitude computed from Equation (\ref{H}).  The statistical uncertainty on $H$ is the same as on $V$ but with an additional systematic uncertainty owing to the unknown phase function}
\tablenotetext{c}{Cross-section in  km$^2$ computed from $H$ using Equation (\ref{area}) with $p_V$ = 0.1}

\end{deluxetable}

\clearpage

\begin{deluxetable}{lcccccccccc}
\tablecaption{Color Measurements
\label{colors}}
\tablewidth{0pt}
\tablehead{
\colhead{Object}  & Date  & Aper\tablenotemark{a}
&  \colhead{B-V}&  \colhead{V-R}&  \colhead{R-I} &  \colhead{B-R}}

\startdata
Borisov & September 15   & 7,500 & -- & 0.49$\pm$0.04 & -- & -- \\
Borisov & September 15   & 15,000 & -- & 0.53$\pm$0.04 & -- & -- \\
Borisov & September 26	& 7,500  & 0.78$\pm$0.05	 & 0.49$\pm$0.03	& 0.49$\pm$0.05	& 1.27$\pm$0.06  \\

Borisov & September 26	& 15,000  & 0.80$\pm$0.05	 & 0.47$\pm$0.03	& 0.49$\pm$0.05	& 1.27$\pm$0.06  \\
\hline
'Oumuamua\tablenotemark{b} & -- &-- & 0.70$\pm$0.06 & 0.45$\pm$0.05 & -- & 1.15$\pm$0.08 \\
Mean LPC\tablenotemark{c} & -- &-- & 0.78$\pm$0.02 & 0.47$\pm$0.02 & 0.42$\pm$0.03 & 1.24$\pm$0.02 \\
Solar Colors\tablenotemark{d}   & & & 0.64$\pm$0.02 & 0.35$\pm$0.01 & 0.33$\pm$0.01 & 0.99$\pm$0.02 \\
\enddata


\tablenotetext{a}{Aperture radius in km}
\tablenotetext{b}{From Jewitt et al.~(2017)}
\tablenotetext{c}{Mean of active long-period comets, 25 observed in B-V, 24 in V-R and 7 in R-I, from Jewitt (2015)}
\tablenotetext{d}{From Holmberg et al.~(2006)}

\end{deluxetable}


\clearpage 

%

\end{document}